\documentclass[aps,prl,twocolumn,showpacs,10pt]{revtex4-1}
\usepackage{graphicx}
\usepackage{amssymb}
\usepackage{amsmath,color}
\usepackage{bbold}
\usepackage{comment}
\pdfoutput=1

\begin{document}

\def\pc{\frac{2\pi}{\Phi_0}}

\def\e{\varepsilon}
\def\f{\varphi}
\def\p{\partial}
\def\ba{\mathbf{a}}
\def\bA{\mathbf{A}}
\def\bb{\mathbf{b}}
\def\bB{\mathbf{B}}
\def\bD{\mathbf{D}}
\def\bd{\mathbf{d}}
\def\be{\mathbf{e}}
\def\bE{\mathbf{E}}
\def\bH{\mathbf{H}}
\def\bj{\mathbf{j}}
\def\bk{\mathbf{k}}
\def\bK{\mathbf{K}}
\def\bM{\mathbf{M}}
\def\bm{\mathbf{m}}
\def\bn{\mathbf{n}}
\def\bq{\mathbf{q}}
\def\bp{\mathbf{p}}
\def\bP{\mathbf{P}}
\def\br{\mathbf{r}}
\def\bR{\mathbf{R}}
\def\bS{\mathbf{S}}
\def\bu{\mathbf{u}}
\def\bv{\mathbf{v}}
\def\bV{\mathbf{V}}
\def\bw{\mathbf{w}}
\def\bx{\mathbf{x}}
\def\by{\mathbf{y}}
\def\bz{\mathbf{z}}
\def\bG{\mathbf{G}}
\def\bW{\mathbf{W}}
\def\Bn{\boldsymbol{\nabla}}
\def\Bo{\boldsymbol{\omega}}
\def\Br{\boldsymbol{\rho}}
\def\Bs{\boldsymbol{\hat{\sigma}}}
\def\bh{{\beta\hbar}}
\def\mA{\mathcal{A}}
\def\mB{\mathcal{B}}
\def\mD{\mathcal{D}}
\def\mF{\mathcal{F}}
\def\mG{\mathcal{G}}
\def\mH{\mathcal{H}}
\def\mI{\mathcal{I}}
\def\mL{\mathcal{L}}
\def\mO{\mathcal{O}}
\def\mP{\mathcal{P}}
\def\mT{\mathcal{T}}
\def\mZ{\mathcal{Z}}
\def\fr{\mathfrak{r}}
\def\ft{\mathfrak{t}}
\newcommand{\rf}[1]{(\ref{#1})}
\newcommand{\al}[1]{\begin{aligned}#1\end{aligned}}
\newcommand{\ar}[2]{\begin{array}{#1}#2\end{array}}
\newcommand{\bra}[1]{\langle{#1}|}
\newcommand{\ket}[1]{|{#1}\rangle}
\newcommand{\av}[1]{\langle{#1}\rangle}
\newcommand{\AV}[1]{\left\langle{#1}\right\rangle}
\newcommand{\aav}[1]{\langle\langle{#1}\rangle\rangle}
\newcommand{\braket}[2]{\langle{#1}|{#2}\rangle}
\newcommand{\ff}[4]{\parbox{#1mm}{\begin{center}\begin{fmfgraph*}(#2,#3)#4\end{fmfgraph*}\end{center}}}
\newcommand{\sh}[1]{\textcolor{red}{#1}}

\title{Superfluid Transport in Quantum Spin Chains}

\author{Silas Hoffman$^1$}
\author{Daniel Loss$^1$}
\author{Yaroslav Tserkovnyak$^2$}
\affiliation{$^1$Department of Physics, University of Basel, Klingelbergstrasse 82, CH-4056 Basel, Switzerland}
\affiliation{$^2$Department of Physics and Astronomy, University of California, Los Angeles, California 90095, USA}

\begin{abstract}
Spin superfluids enable long-distance spin transport through classical ferromagnets by developing topologically stable magnetic textures. For small spins at low dimensions, however, the topological protection suffers from strong quantum fluctuations. We study the remanence of spin superfluidity inherited from the classical magnet by considering the two-terminal spin transport through a finite spin-1/2 magnetic chain with planar exchange. By fermionizing the system, we recast the spin-transport problem in terms of quasiparticle transmission through a superconducting region. We show that the topological underpinnings of a semiclassical spin superfluid relate to the topological superconductivity in the fermionic representation. In particular, we find an efficient spin transmission through the magnetic region of a characteristic resonant length, which can be related to the properties of the boundary Majorana zero modes.
\end{abstract}
 
\maketitle

\textit{Introduction.}--In magnetic insulating materials spin transport is mediated via spin-wave excitations or magnons rather than electrons \cite{khitunIEEE08,meierPRL03}. Because the excitations in ferromagnetic insulators are bosonic, magnons are capable of supporting Bose-Einstein condensates \cite{nikuniPRL00,oosawaJoPCM99,raduPRL05,demokritovNAT06,nakataPRB14,nakataJoPD17} and even spin superfluid transport \cite{soninJETP78,soninAP2010,konigPRL01,benderPRL12,chenPRB14,chenPRB14ssf,chenPRL15,takeiPRL14,takeiPRB14}.

For a quasi-one-dimensional easy-plane magnet, the magnetic order is topologically characterized by the winding number of the mapping from $\mathbb{R}^1$ to $S^1$. When a spin bias is applied to the boundary of such a system, topological defects in the magnetic texture, which are characterized by nontrivial winding numbers, are nucleated \cite{kimPRB16b}. The ensuing topological transport yields a long-range spin supercurrent \cite{takeiPRL14,kimPRB15} subject to thermal \cite{kimPRB16} or quantum \cite{kimPRL16} phase slips. Such a supercurrent is suppressed, however, when the topological protection is destroyed by applying a magnetic field greater than the in-plane anisotropy. A preferred (easy) axis within the plane, furthermore, can reduce the mobility of the topological texture \cite{kimPRB15}. 

In contrast to (semi)classical magnets, the elementary excitations in quantum spin chains exhibit strong quantum fluctuations. In particular, in the extreme case of the lowest spin 1/2, it is unclear to which extent the superfluid character of the winding dynamics is applicable and useful. Recent spin-caloritronic experiments on spin liquids have demonstrated that spin can be transported via quantum spin excitations by thermal biasing \cite{hirobeNAT16}. With these practical tools in hand, an important open question concerns the possibility of long-ranged collective spin flows in quantum spin chains.

\begin{figure}[h]
\includegraphics[width=1\linewidth]{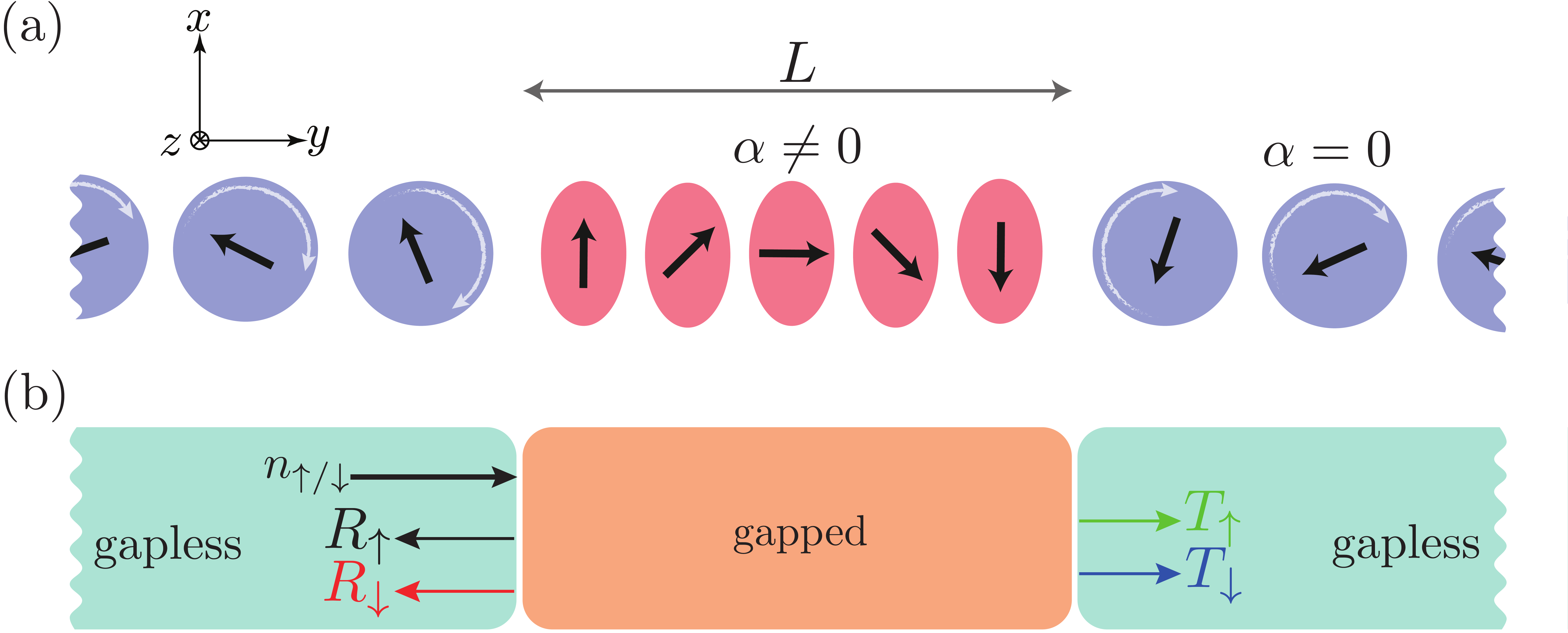}
\caption{ (a) Schematic of our spin-1/2 chain setup: The left and right sides are semi-infinite spin chains (blue circles) wherein the spins are symmetrically coupled in the $xy$ plane via an exchange coupling. The central region is of length $L$ and has an in-plane anisotropy parameterized by $\alpha$. (b) The left and right sides, absent of anisotropy, have a gapless spectrum while the anisotropic central region is gapped. An incoming spin excitation, which is generically in a superposition of positive ($n_\uparrow$) and negative  ($n_\downarrow$) spin collinear with the $z$ axis, can be reflected (transmitted) as a spin up, $R_\uparrow$ ($T_\uparrow$), or spin down, $R_\downarrow$ ($T_\downarrow$), excitation.
}
\label{setup}
\end{figure}

In this Letter, we consider two semi-infinite XY spin chains, which realize Fermi-liquid like spin reservoirs (Fig.~\ref{setup}). They supply and drain spin currents from a central region, whose transport is examined with an eye on spin superfluidity. We control the spin ordering and, consequently, the transport properties of the central region by applying an out-of-plane magnetic field, which, in the semiclassical view, would tune the superfluid density, and an axial anisotropy within the easy ($xy$) plane, which breaks rotational symmetry and would pin the condensate phase. When the spins are uniformly ordered by a sufficiently large magnetic field, transport of low energy excitations between the reservoirs is exponentially suppressed with the length of the central region. 
A chain with an easy-plane anisotropy and a sufficiently small applied magnetic field affords zero-energy excitations which are transported ballistically. Although the bulk spectrum is gapped when there is an easy-axis anisotropy in the $xy$ plane, evanescent domain walls at the ends of the chain survive which contribute to the transport. This is explicated by performing a nonlocal transformation which maps the spin operators to fermions. In the fermionic language, the localized domain walls correspond to Majorana end modes. Analogous to the effect Majoranas have on the charge transport in topological superconductors, these localized domain walls qualitatively affect the transport in anisotropic spin chains. Specifically, for a sufficiently long central region, zero-energy excitations carrying positive spin along the $z$ axis are perfectly reflected off the central region carrying negative spin; this is the analogue of perfect Andreev reflection from a topological superconductor \cite{lawPRL09}. Furthermore, zero-energy excitations can be ballistically transported through the central region when it is a certain resonant length, defined below, tunable by an applied magnetic field. This corresponds to perfect conductance of a fermion through a topological superconductor of the same resonant length.

\textit{Model.--}A simple model to illustrate quantum transport is an $N$-site spin-$1/2$ ferromagnetic chain with a planar exchange coupling
\begin{equation}
H=-J\sum_{i=1}^{N-1}\left[(1+\alpha)\sigma_i^x\sigma_{i+1}^x+(1-\alpha)\sigma_i^y\sigma_{i+1}^y\right]-h\sum_{i=1}^{N}\sigma_i^z\,,
\label{h_XY}
\end{equation}
where $\sigma^\mu_i$ for  $\mu=x,y,z$ are the Pauli matrices acting on a spin at site $i$. Here, $J$ is the exchange coupling between adjacent sites, $\alpha$ parameterizes the asymmetry in the $xy$ plane, and $h$ is the magnitude of an applied magnetic field along the $z$ axis. Lengths are measured in units of the lattice spacing $a$. For the following discussion, we assume ferromagnetic exchange and so restrict the parameters as such, $J>0$ and $0\leq\alpha\leq1$ \footnote{In the case of an antiferromagnet exchange, $J<0$, the analysis formally remains the same with $k\rightarrow k+\pi$.}. If there is no anisotropy, $\alpha=0$, the Hamiltonian is rotationally symmetric about the $z$ axis. For finite $\alpha$, this symmetry is reduced to rotations by $\pi$. There is a quantum phase transition when the magnetic field is equal to the exchange, $|h|=J$. When $|h|>J$, the ground state is a nondegenerate paramagnet in which the spins align according to the sign of the magnetic field. When $|h|<J$ and in the absence of anisotropy, $\alpha=0$, the ground state is symmetric under rotations about the $z$ axis. For a finite anisotropy, $\alpha\neq0$, this symmetry is reduced to rotations by $\pi$ and the ground state is doubly degenerate.

\begin{figure}[h]
\includegraphics[width=1\linewidth]{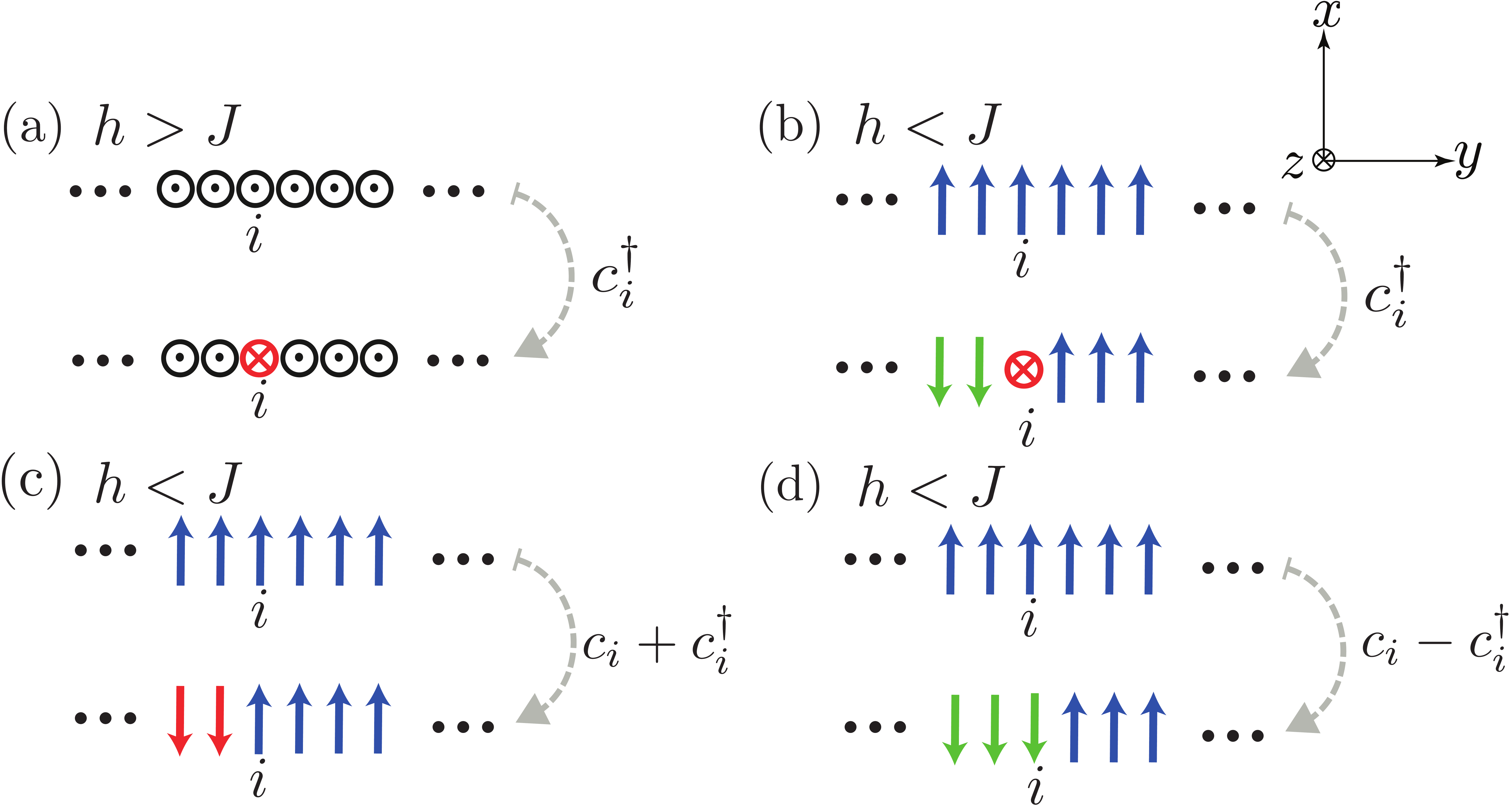}
\caption{The action of a fermionic creation operator, $c_i^\dagger$, (a) flips the spin at site $i$ when $|h|>J$  and (b) creates a domain wall pointing along the $z$ axis when acting on the degenerate ground state, for $|h|<J$. (c) A superposition of $c_i+c_i^\dagger$  [(d) $c_i-c_i^\dagger$] rotates all sites before $i$ by $\pi$ around the $z$ axis and the site $i$ by $\pi$ around the $x$ ($y$) axis. 
}
\label{excite}
\end{figure}

The spectrum is found upon performing a Jordan-Wigner transformation \cite{liebAoP61,tserkovnyakPRA11}. Defining a spinless fermionic creation (annihilation) operator at site $j$, $c_j=\sigma^-_j \mathcal P_j$ ($c_j^\dagger=\sigma^+_j\mathcal P_j$) where $\sigma_j^\pm=(\sigma_j^x\pm i\sigma_j^y)/2$ and $\mathcal P_j=\prod_{l<j}(-\sigma_l^z)$. That is, $c_j^\dagger$ or $c_j$ polarize the spin at site $j$ parallel or antiparallel to the $z$ axis, respectively, while the sites before $j$ are rotated by $\pi$ around the $z$ axis. This corresponds to a spin flip at site $j$ when acting on the paramagnetic ground state [Fig.~\ref{excite}(a)]. In the doubly degenerate phase, the excitation is a domain wall at site $j$ which is polarized parallel or antiparallel to the $z$ axis  [Fig.~\ref{excite}(b)] and is the ferromagnetic analogue of the Villain mode \cite{braunIJMPB96}. Using these fermionic operators, Eq.~(\ref{h_XY}) becomes
\begin{equation}
H=-\frac{J}{2}\sum_{i=1}^{N-1}\left(c_i^\dagger c_{i+1}+\alpha c_i^\dagger c_{i+1}^\dagger+\textrm{H.c}\right)-h\sum_{i=1}^{N}( c_i^\dagger c_i -1/2)\,.
\label{kitaev}
\end{equation}
This is the Kitaev chain \cite{kitaevPU01}, describing a spinless metal ($p$-wave superconductor) for $\alpha=0$ ($\alpha\neq0$). The bulk excitations are known \cite{aliceaNATP11} and can be found in terms of the Fourier-transformed operators $c_k$ and $c_k^\dagger$ \cite{supp_qss}. In the spin chain (metal) picture, $c_k$ creates holes carrying $-\hbar$ spin quantized along the $z$ axis (negative charge) while $c_k^\dagger$ creates particles carrying $\hbar$ spin (positive charge). 

When the magnetic field is small, $h<J$, and the chain is absent of anisotropy, $\alpha=0$, the spectrum consists of a partially filled gapless band \cite{aliceaNATP11}. There are two zero-energy modes with $\pm k_0=\pm\cos^{-1}(h/J)$. A superposition of $c_{k_0}$ and $c_{k_0}^\dagger$ creates an excitation which changes spin direction in the $xy$ plane throughout the chain:
\begin{equation}
e^{i\varphi}c_{k_0}+ e^{-i\varphi}c_{k_0}^\dagger=\sum_j\mathcal P_j[\cos(k_0 j+\varphi)\sigma_j^x+\sin(k_0 j+\varphi)\sigma_j^y]\,,
\label{dom}
\end{equation}
where $\varphi$ is an arbitrary phase. That is, suppose this operator acts on a chain uniformly polarized in the $xy$ plane. The resultant state is a delocalized domain wall that rotates clockwise with wavelength $1/2k_0$, creating a spiral in the magnetic texture \cite{popkovPRE13,popkovPRA16,popkovPRE17,popkovPRA17,posskeCM18}. Similarly, taking $k_0\rightarrow-k_0$, the state is a domain wall rotating counter-clockwise. Note that these are delocalized Majorana fermions as they are Hermitian.

In the doubly degenerate ground state, $|h|<J$ and $\alpha\neq0$, the system is no longer rotationally invariant and the bulk zero-energy modes are gapped out. However, in a finite or semi-infinite chain, there exist zero-energy modes at the ends. In the fermionic language, these are the celebrated Majorana zero modes \cite{kitaevPU01} while, in the spin language, they are localized zero-energy domain walls \cite{tserkovnyakPRA11}. Together, these end modes form a nonlocal complex fermionic state which can be occupied or unoccupied, parameterizing the double degeneracy of the ground state. We focus on the regime when $h<0$ and $|h|\lesssim J$, so that the band is nearly depleted which corresponds to the spin chain largely polarized antiparallel to the $z$ axis \footnote{The discussion follows identically when $h>0$, upon exchanging spins parallel to the $z$ axis and spins antiparallel to the $z$ axis or, in the fermionic language, exchanging particles and holes and expanding around $k=\pi$ rather than $k=0$.}. Because $k\ll1$, we can pass from a discrete coordinate to a continuum, $\ell$, and the mode at the right end is created by the operator $\int d\ell  \sigma_\ell^x \mathcal P_\ell(e^{-\kappa^+\ell}-e^{-\kappa^-\ell})$. The mode at the left end of a semi-infinite chain is, similarly, created by $\int d\ell \mathcal  \sigma_\ell^y \mathcal P_\ell(e^{ \kappa^+\ell}-e^{\kappa^-\ell})$. These right and left modes correspond to exponentially localized, on the length scale $1/\textrm{Re}[\kappa^\pm]$, zero-energy domain walls that are created by a $\pi$ rotation around the $x$ and $y$ axis, respectively. When $|h|>J$, the system is trivially gapped and there exist no zero-energy bulk or localized modes. 

\textit{Transport.--}To calculate the transport properties of a finite size chain, consider a geometry in which the translational symmetry is broken: two semi-infinite isotropic spin chains ($\alpha=0$) are connected to either side of a finite anisotropic chain ($\alpha\neq0$) of length $L$. See Fig.~\ref{setup}(a). The left and right isotropic sections of the chain are leads which provide a gapless source and drain of spin excitations, respectively, which probe the transport properties of the central gapped anisotropic region. Our setup is equivalent to a normal metal$|p$-wave superconductor$|$normal metal junction through which charge transport is mapped to spin transport in the spin chain [Fig.~\ref{setup}(b)]. In general, the leads are held at a different magnetic field, $h'$, from the magnetic field of the central region, $h$. Furthermore, to enter the central region, suppose excitations must overcome an energy barrier $U$ such that when $U=0$ the leads are open and when $U/J\rightarrow\infty$, the central region is totally disconnected \footnote{Specifically, after going from the discrete chain to the continuum limit, we include a delta-function potential of strength $U$ separating the leads from central region. Physically, this corresponds to local a magnetic field (chemical potential), on the scale of the lattice spacing, in the spin (electronic) picture. This barrier does not qualitatively affect our results but, for large $U$, makes the effects more dramatic.}.

In the following we focus on the continuum limit of the system and proceed to calculate the scattering amplitudes in the fermionic description by matching the solutions at the interfaces between the leads and central region. Consider a right-moving excitation in the left lead with energy $E$. In general, this can be a superposition of a particle carrying positive spin with wave vector $k^>=\sqrt{1+(h'+ E)/J}$ and a hole carrying negative spin with wave vector $k^<=\sqrt{1+(h'- E)/J}$. The weight of the particle (hole) in the wave function is parameterized by $n_\uparrow$ ($n_\downarrow$). Because spin along the $z$ axis is not conserved in the central region, 
the incoming excitation can be reflected as a particle or a hole with probability $R_\uparrow$ or $R_\downarrow$, respectively. The excitation can likewise be transmitted to the right lead as a particle (hole) with probability $T_\uparrow$  ($T_\downarrow$). 

Consider the regime near the topological phase transition, $\alpha^2\gg |1+h/J|$, in which the spectrum is gapped by $|1+h/J|$ at $k=0$ \cite{supp_qss}. First, this limit allows us to contrast the transport properties in the degenerate, $|h|<J$, and nondegenerate, $|h|>J$, phases with equal gaps. Second, zero-energy in-gap states have two decay lengths which are well-separated  $1/\kappa^+=1/\alpha\ll\kappa^-=\alpha/(1+h/J)$ and allow us to obtain simple analytic solutions for the transport properties when $L\sim1/\kappa^-$.

\begin{figure}[h]
\includegraphics[width=1\linewidth]{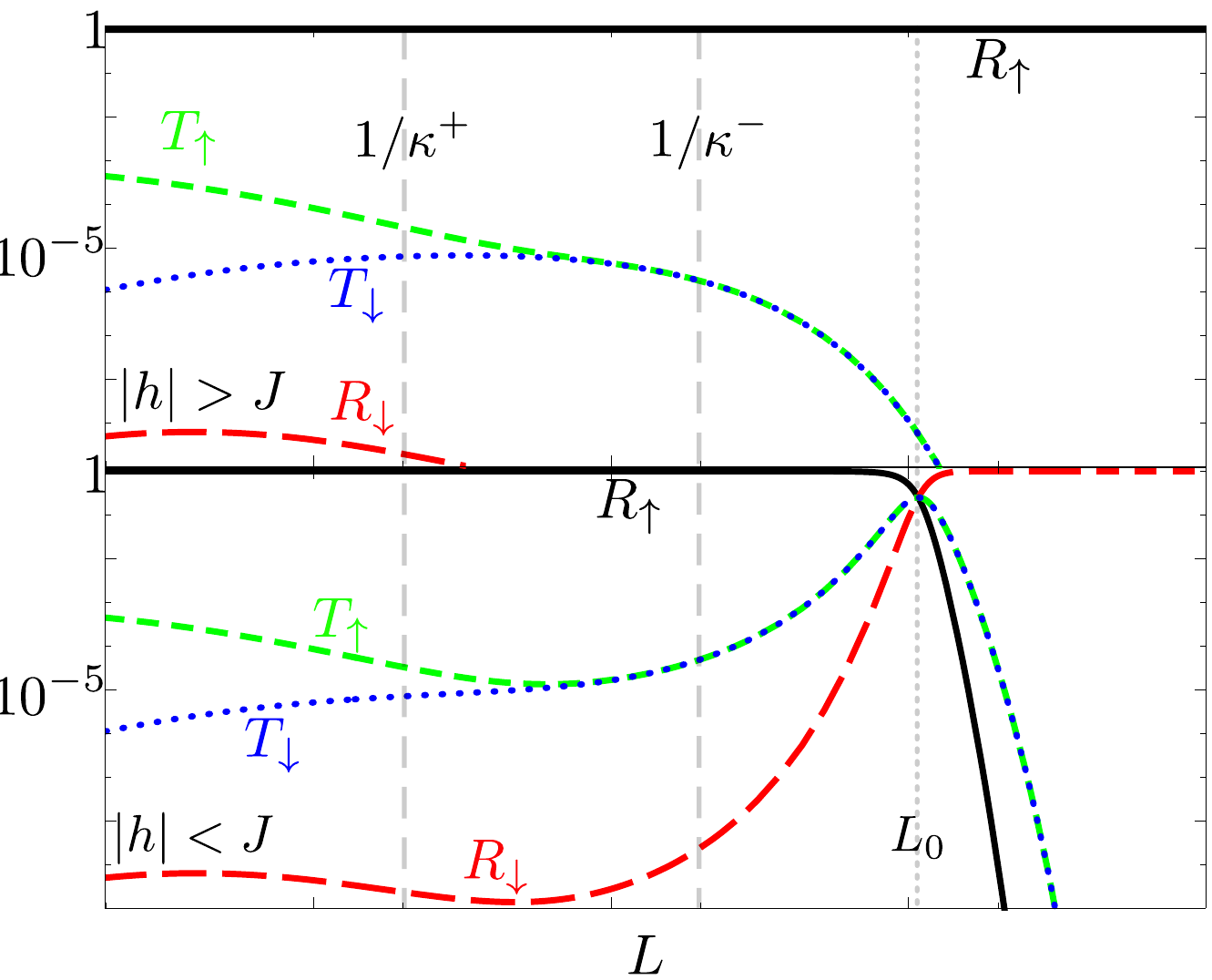}
\caption{The probability as a function of $L$ of a zero-energy $\hbar$-spin excitation impinging on an ordered spin chain of length $L$ to be reflected, $R_\uparrow$ ($R_\downarrow$), with the same (opposite) spin or to be transmitted carrying positive (negative) spin, $T_\uparrow$ ($T_\downarrow$). The plots are logarithmic on both the axes and $\alpha=1$, $U/J=10$, and $h'/J=-0.9$. The upper panel is in the nondegenerate phase, $h/J=-1.1$, while the lower panel is in the degenerate phase, $h/J=-0.9$.
}
\label{trans}
\end{figure}

We study a zero-energy excitation with spin along the $z$ axis impinging on the central region, $E=0$ and $n_\uparrow=1$. When the length of the central region is short, $\kappa^+L\lesssim1$, the transport properties of both degenerate and nondegenerate phases are characterized by an exponential suppression of the transmission and perfect reflection (Fig.~\ref{trans}). For $\kappa^+L\gg1$, the two phases show a qualitative difference. In the nondegenerate phase, the transmittance remains exponentially suppressed and the reflection is perfect [Fig.~\ref{trans} (upper panel)]. In the degenerate phase, the transmission and reflection probabilities are \cite{supp_qss}
\begin{align}
T_\uparrow&=T_\downarrow=\textrm{sech}^2[\kappa^-(L-L_0)]/4\,,\nonumber\\
R_\uparrow&=e^{-2\kappa^-(L-L_0)}\textrm{sech}^2[\kappa^-(L-L_0)]/4\,,\nonumber\\
R_\downarrow&=e^{2\kappa^-(L-L_0)}\textrm{sech}^2[\kappa^-(L-L_0)]/4\,.
\label{transp}
\end{align}
where the resonant length, 
\begin{equation}
L_0= \frac{\alpha}{1+h/J}\ln\left[\frac{1+h'/J+
(\alpha/4 + U/J)^2}{\alpha \sqrt{1+h'/J} }\right]\,.
\label{res}
\end{equation}
When $1/\kappa^-\lesssim L < L_0$, the probability of transmission increases exponentially as the length of the central region increases [Eq.~(\ref{transp}) and Fig.~\ref{trans} (lower panel)]. At $L=L_0$, the probability to transmit a zero energy excitation is locally maximized and $T_\uparrow=T_\downarrow=R_\uparrow=R_\downarrow=1/4$. Beyond $L_0$, the transmission is exponentially suppressed and the particle is  favored to be reflected as a hole. That is, a spin of $2\hbar$ is perfectly injected into the anisotropic region. This is the spin chain analogue of perfect Andreev reflection in one dimensional topological superconductors \cite{lawPRL09}. Because $L_0$ is inversely proportional to $h$ [Eq.~(\ref{res})], for a fixed $L$, the probability to transmit the particle can be tuned by changing $L_0$ according to the applied magnetic field. 

When $|h|>J$, transmission is suppressed for all $L$ because there exist no in-gap states states in the nondegnerate phase. In the degenerate phase, on the other hand, there exist in-gap evanescent end states which can enhance transport. When $L\gg L_0$, the states do not overlap and there is no coherent transmission of the signal between the leads. Precisely at $L=L_0$, the end modes overlap but are stabilized at zero energy because they leak into the leads. For $U\gg J$, the operator creating such a zero-energy state is
\begin{align}
\int d\ell\mathcal P_\ell&\left[\cos(\pi\ell/2L_0)\sigma_\ell^x (e^{-\kappa^+\ell}-e^{-\kappa^-\ell})\right.\nonumber\\
&\left.+\sin(\pi\ell/2L_0)\sigma_\ell^y (e^{\kappa^+\ell}-e^{\kappa^-\ell})\right]\,,
\label{maj}
\end{align}
which interpolates  between the left and right zero-energy modes. Physically, this corresponds to a delocalized domain wall, exponentially weighted on the left and right ends, whose direction changes continuously by $\pi$ between the two ends. Because $T_\uparrow=T_\downarrow$, no net spin is transferred between the leads. Nonetheless, a spinless flux of excitations should induce correlations between the leads and manifest in the spin noise. 

When $L<L_0$, the end modes overlap and hybridize away from zero energy. Suppose an in-gap but finite energy, $E$, excitation impinges on the central region. There is a peak in the transmission probability at a different resonant length which is smaller than $L_0$ \cite{supp_qss}. Furthermore, for an energy near the gap edge, $E\lesssim |h+J|$, the transmission probability, as a function of length, has the form of a Lorentzian rather than exponential as in Eq.~(\ref{transp}). Because the mode interpolating between the leads is not at zero energy, the probability of transmission for a positive spin is different than for a negative spin, resulting in a net flow of spin. This restoration of long distance transmission of spin is the remanence of classical spin supercurrent in the ordered quantum spin chain.

According to Eq.~(\ref{maj}) these zero-energy domain walls in the central region are created by a rotation around an axis in $xy$ plane. One may suspect that there would be an increase in transmission if the incoming excitation from the source was also created by an in-plane rotation, \textit{i.e.} $|n_\uparrow|=|n_\downarrow|=1/\sqrt{2}$, in the sense of Eq.~(\ref{dom}). Indeed, we find that when $n_\uparrow e^{-i\chi}=n_\downarrow=1/\sqrt{2} $ for 
\begin{equation}
\chi=\tan^{-1}\left[\frac{4 \sqrt{1+h'/J} (\alpha+ 2 U/J)}{1+h'/J-(\alpha/2+U/J)^2 )}\right]\,,
\end{equation} 
the transmission is perfect; the probability for an excitation to be transmitted to the right lead with positive spin and negative spin are equal and their sum is one. When the relative phase between the components of the excitation is $\chi+\pi$, the transmission probability is zero. Because an incoming excitation with spin collinear with the $z$ axis can be expressed as an equal superposition of in-plane zero-energy excitations \footnote{Taking Eq.~(\ref{dom}) with $\varphi=0$ ($\varphi=\pi/2$), one obtains $c_{k_0}+c_{k_0}^\dagger$ ($ic_{k_0}-ic_{k_0}^\dagger$). The sum or difference of these terms, with a relative factor of $i$, garners $c_{k_0}$ or $c_{k_0}^\dagger$, respectively, up to a constant.}, maximally, half is transmitted while the other half is reflected, consistent with Eq.~(\ref{transp}). 

\textit{Discussion.--}  We find that zero-energy excitations can be transported through degenerate gapped quantum spin chains at a resonant length, $L_0$, but the transmission probability is exponentially suppressed away from this length. This is due to the long-range order of the spins along the axis of anisotropy. In contrast, the leads are isotropic in the $xy$ plane and lack order, allowing zero-energy modes to propagate ballistically. An easy axis gaps out these bulk modes. Semiclassically, topological defects can tunnel through this barrier or, upon applying a sufficiently large spin bias, overcome it energetically. Because evanescent modes are present only in the quantum regime, it is a feature unique to the quantum system that low energy excitations can persist over long distances thereby partially restoring the superfluidity \cite{ochoaPRB18}.

Suppose that the chain is made up of several elements each with a random in-plane anisotropy, as result of defects for instance, which locally order the chain. Adjacent elements whose anisotropy differs by an angle $\phi$ are equivalent to superconducting weak links in a Kitaev chain. Such a topological Josephson junction can support in-gap evanescent states whose energy is proportional to $\sin\phi$ where $\phi$ is \textit{half} of the difference in phase across the junction \cite{kitaevPU01}. Hybridization of these localized states can form an in-gap band capable of supporting spin excitations. That is, disorder in an easy-axis spin chain can globally destroy the order on average, thereby restoring low-energy ballistic spin transport.

Throughout this manuscript, we have neglected out-of-plane exchange interactions, \textit{i.e.} along the $z$ axis, between neighboring sites. It is known that such an \textit{antiferromagnetic} exchange, corresponding to a repulsive interaction in the fermionic picture, can modify the order \cite{lossPRL95} and destroy the end states \cite{gangPRL11} for sufficiently large interaction. As a result, excitations can be perfectly reflected at the interface with the anisotropic region even when the magnetic field is smaller than the exchange \cite{fidkowskiPRB12}. Upon the inclusion of an out-of-plane \textit{ferromagnetic} exchange interaction on the other hand, which corresponds to an attractive interaction in the fermionic picture, perfect spin injection into the anisotropic region remains even for a large out-of-plane exchange and can persist for large applied magnetic fields \cite{fidkowskiPRB12}. We leave the mechanism supporting this property and the length dependence of the transport \cite{aseevCM18} as an open question for future work.

\textit{Acknowledgements.--}  The authors would like to acknowledge beneficial discussions with Victor Chua. This work was supported by the Swiss NSF, NCCR QSIT, and NSF under Grant No. DMR-1742928.

\newpage

\onecolumngrid

\bigskip

\begin{center}
\large{\bf Supplemental Material to `Superfluid Transport in Quantum Spin Chains' \\}
\end{center}
\begin{center}
Silas Hoffman$^1$, Daniel Loss$^1$, and Yaroslav Tserkovnyak$^2$\\
{\it $^1$Department of Physics, University of Basel, Klingelbergstrasse 82, CH-4056 Basel, Switzerland}
\\{\it $^2$Department of Physics and Astronomy, University of California, Los Angeles, California 90095, USA}
\end{center}
\twocolumngrid
\section{Kitaev Hamiltonian}
\begin{figure}[h]
\includegraphics[width=1\linewidth]{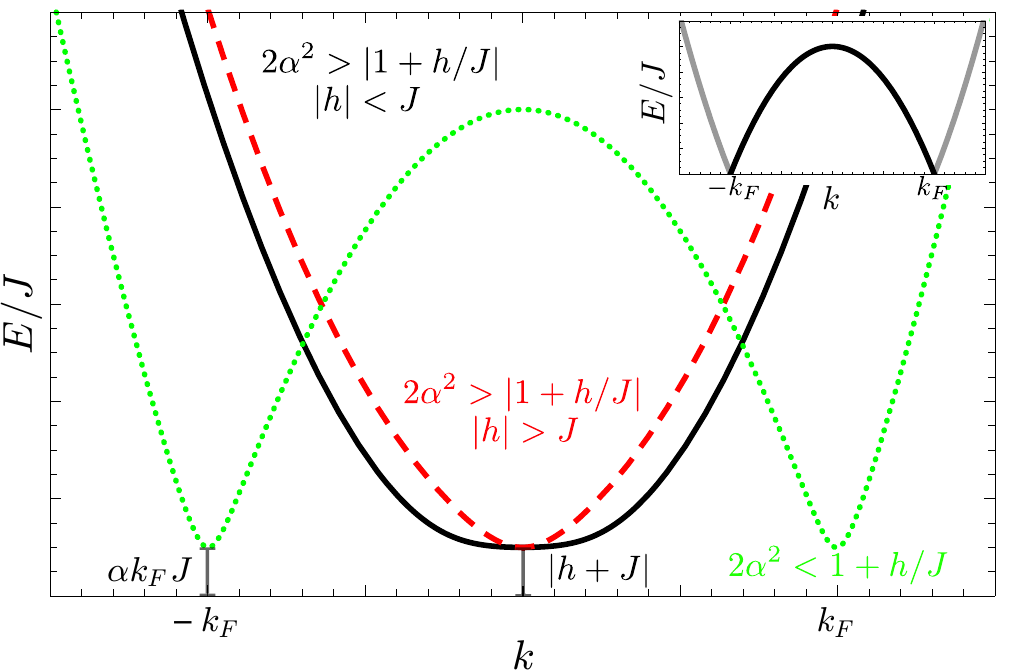}
\caption{The positive energy spectrum of the spin chain system with periodic boundary conditions in three regimes: (1)  $2\alpha^2>|1+h/J|$ and $|h|<J$ (black solid curve), (2)  $2\alpha^2>|1+h/J|$ and $|h|>J$, and (3) $2\alpha^2<1+h/J$. Inset: The positive energy spectrum when $\alpha=0$ and $|h|<J$. The excitations for $k<|k_F|$ ($k>|k_F|$) correspond to particles (holes).}
\label{spec}
\end{figure}

The Kitaev Hamiltonian, given by Eq.~(2) in the main text, can be Fourier transformed to momentum space taking the form $H=\frac{1}{2}\sum_k C_k^\dagger \mathcal H C_k$, where the sum is over $k$ in the Brillouin zone and 
\begin{equation}
\mathcal H =-(J\cos k +h)\eta_z-(\alpha J\sin k )\eta_y\,,\,\,\,\,\, C^\dagger_k=[c_k^\dagger,c_{-k}]\,.
\label{kitC}
\end{equation}
Here, $\eta_j$ are the Pauli matrices acting in Nambu space. In the following we are interested in long wavelengths as compared to the lattice spacing, which is valid when $|h|$ is comparable to $J$, so that the low energy Hamiltonian is
\begin{equation}
\mathcal H =J[ k^2-(1+h/J)]\eta_z-k \alpha J \eta_y\,,
\end{equation}
where we henceforth take $h<0$. When $\alpha=0$, the energies are $\pm J[k^2-(1+h/J)]$ for the respective eigenvectors $\varphi^+=(1,0)$ and $\varphi^-=(0,1)$. The spectrum has two Fermi points, $\pm k\pm\sqrt{1+h/J}$ [Fig.~\ref{spec} (inset)]. If $|h|>J|$ the spectrum has a gap of $|h+J$ at $k=0$. When $\alpha\neq0$, the eigenvalues and eigenvectors are given by
\begin{align}
E^\pm/J&=\pm\sqrt{[k^2-(1+h/J)]^2+\alpha^2k^2}\,,\nonumber\\
\phi_k^\pm&=\left[\frac{k^2-(1+h/J)+E^\pm/J}{-i\alpha k},1\right]\,,
\label{sol_an}
\end{align}
respectively. 

For finite $\alpha$, the spectrum $E^\pm$ has a gap which closes at $k=0$ when $|h|=J$ signaling a phase transition with $|h|<J$ ($|h|>J$) supporting a degenerate (nondegenerate) ground state. There are two qualitatively different regimes of the spectrum: (1) when $2\alpha^2>|1+h/J|$ [Fig.~\ref{spec} (black solid and red dashed curve)] and (2) when $2\alpha^2<(1+h/J)$ [Fig.~\ref{spec} (green dotted curve)]. In the first case, there is one minimum in the spectrum at $k=0$ with gap $|h+J|$. Near the phase transition when the energy is within the gap, $\alpha J\gg|h+J|>E$, all the wave vectors are purely imaginary, given by $\pm i \alpha$ and $\pm i \sqrt{(h+J)^2-E^2}/\alpha J$, \textit{i.e.} there are no propagating solutions. When the energy is above the gap but still near the bottom of the band ($\alpha J\gg E>|h+J|$), there are two propagating solutions, $\pm \sqrt{E^2-(h+J)^2}/\alpha J$, and two totally imaginary wave vectors, $\pm i \alpha$. In the second case there are two minima in the spectrum which are symmetric about $k=0$ where there is a local maximum. Deep within the degenerate regime, $(1+h/J)\gg\alpha^2$, the minima are at $\pm k_F$ with gap $\alpha k_F J$. When the energy is within the gap, $E<\alpha k_F J$, the four wave vectors are $k_F\pm i\sqrt{\alpha^2-E^2/J(h+J)}$ and $-k_F\pm i\sqrt{\alpha^2-E^2/J(h+J)}$; they oscillate with wave vector $k_F$ and decay or grow exponentially according to their depth within the gap. Above the gap with $E<\sqrt{(h+J)^2+J(h+J)\alpha^2}$, there are four propagating states with $k_F\pm \sqrt{E^2/J(h+J)-\alpha^2}$ and $-k_F\pm \sqrt{E^2/J(h+J)-\alpha^2}$, \textit{i.e.} two solutions around each Fermi point. When $E>\sqrt{(h+J)^2+J(h+J)\alpha^2}$, there are two purely imaginary and two purely real solutions symmetric about $k=0$.

\section{Analytic Form of Transport Coefficients}

In general the solutions for the transport coefficients are rather complicated. However, when $|1+h/J|\ll1$, the separation in length scales allows us to obtain an analytic formula for these coefficients in two regimes: $\kappa^-L\ll 1$ and $\kappa^+L\gg 1$. Because we are interested in the long length behavior in the degenerate phase, we focus on the regime when $L\gg \kappa^+$ and $|h|<J$.
\begin{widetext}
\subsection{Zero energy}

When the incoming excitation is at zero energy, $E=0$, the transport coefficients are
\begin{align}
t_\uparrow&=4 i \alpha  k_F e^{(-i k_F - \kappa^-)L}\frac{[(\alpha + 2 U/J)^2 + (2 k_F)^2 ]n_\downarrow-(\alpha + 2 i k_F + 2 U/J )^2 n_\uparrow}{[(\alpha + 2 U/J )^2 + (2 k_F)^2 ]^2  e^{-2\kappa^- L} + (4\alpha k_F )^2}\,,\nonumber\\
t_\downarrow&=4 i \alpha  k_F e^{(i k_F - \kappa^-)L}\frac{(\alpha - 2 i k_F + 2 U/J )^2n_\downarrow- [(\alpha + 2 U/J)^2 + (2 k_F)^2 ]n_\uparrow}{[(\alpha + 2 U/J )^2 + (2 k_F)^2 ]^2  e^{-2\kappa^- L} + (4\alpha k_F)^2}\,,\nonumber\\
r_\uparrow&=-\frac{ [(\alpha - 2 i k_F + 2 U/J) (\alpha + 2 i k_F + 2 U/J)^3]n_\uparrow e^{-2\kappa^- L}+(4\alpha k_F)^2  n_\downarrow}{[(\alpha + 2 U/J )^2 + (2 k_F)^2 ]^2  e^{-2\kappa^- L} + (4\alpha k_F )^2}\,,\nonumber\\
r_\downarrow&=-\frac{ [(\alpha - 2 i k_F + 2 U/J) (\alpha + 2 i k_F + 2 U/J)^3] n_\downarrow e^{-2\kappa^- L}+(4\alpha k_F)^2n_\uparrow}{[(\alpha + 2 U/J )^2 + (2 k_F)^2 ]^2  e^{-2\kappa^- L} + (4\alpha k_F )^2}\,.
\end{align}
where we have redefined $k_F=\sqrt{1+h'/J}$ to be the Fermi points in the leads.

The complex conjugate square of these quantities are the transport probabilities in the main text: $R_\uparrow=|r_\uparrow|^2$, $R_\downarrow=|r_\downarrow|^2$, $T_\uparrow=|t_\uparrow|^2$, and $T_\downarrow=|t_\downarrow|^2$. One can show that the denominator of the transmission is minimized for the resonant length, $L_0$ [Eq.~(5) in the main text] which is independent of the polarization of the incoming excitation.

When the magnitude of the incoming spin up and down excitation is equal but differ in a phase, $n_\uparrow=n_\downarrow e^{i\chi}$,

\begin{align}
|t_\uparrow|&=|t_\downarrow|=\frac{\sqrt{2}4 \alpha  k_F \sqrt{(\alpha + 2 U/J)^2 + (2 k_F)^2 } [2 k_F \cos(\chi/2) + (\alpha + 2 U/J) \sin(\chi/2)] e^{ \kappa^-L}}{[(\alpha + 2 U/J )^2 + (2 k_F)^2 ]^2  + (4\alpha k_F)^2 e^{2\kappa^- L} }\,.
\end{align}

One can show that the transmission is maximized when
\begin{equation}
e^{i\chi}=-\frac{\alpha - 2 i k_F + 2 U/J }{\alpha + 2 i k_F + 2 U/J}\,,
\end{equation}
which is equivalent to Eq.~(7) in the main text. Using the condition $\alpha^2\gg|1+h/J|$, when $U=0$ (
$U/J\gg\alpha$) we find $\chi\approx0$ ($\chi\approx\pi$).

\subsection{Finite energy}
We now consider the transmission coefficients of an excitation with positive energy within the gap, $0<E<h+J$, scattering off the central region. A simple form of the transport coefficients can be found as the energy approaches the band edge $E\rightarrow h+J$,
\begin{align}
t_\uparrow&= e^{-i k_F L} \frac{w_\uparrow}{u+ v L}\,,\,\,\,\,\,t_\downarrow= e^{-i k_F L} \frac{w_\downarrow}{u+ v L}\,,\nonumber\\
w_\uparrow&=4 \alpha k_F J^3 [-i \alpha J (n_\uparrow - n_\downarrow) + 2 k_F J (n_\uparrow + n_\downarrow) - 2 i U (n_\uparrow - n_\downarrow) ] [\alpha^3 J^2 + 2 \alpha^2 J (i k_F  J+ U) + 4 (h+J)(i k_F J + U)]\,,\nonumber\\
w_\downarrow&=4\alpha^2 k_F J^3[-i \alpha J (n_\uparrow - n_\downarrow) + 2 k_F J (n_\uparrow + n_\downarrow) - 2 i U(n_\uparrow - n_\downarrow) ] [\alpha^2 J^2 - 2 J (h+J) + 
   2 \alpha J (-i k_F J+ U)]\,,\nonumber\\
u&=\{\alpha^2 J^2 + 4 \alpha J (i k_F J+ U) + 4 [(k_F J)^2 + U^2]\} \nonumber\\
&~~\times\{\alpha^4 J^4 + 4 \alpha^3 J^3 (-i k_F J + U) + 
   8 J (h+J) [(k_FJ)^2 + U^2] + \alpha^2 J^2 [4(k_FJ)^2 - 2 J (h+J)  + 4 U^2]\}\,,\\
v&=-\alpha J (h+J) \{\alpha^2 J^2 + 4 \alpha J (i k_FJ + U) + 4 [(k_FJ)^2 + U^2]\}^2\,.
\end{align}
Note that to obtain this expression we have assumed that the energy of the excitation is much smaller than $h'+J$.

In general, $|t_\uparrow|^2$ and $|t_\downarrow|^2$ are unnormalized Lorentzian functions of $L$ whose prefactor,  center, and width are complicated functions of the system parameters. To further simplify the expressions, consider the case of when the excitation in the left lead carries $\hbar$ spin, $n_\uparrow=1$ and $n_\downarrow=0$. When $U$ is large and making use of the limit $\alpha\gg(1+h/J)$, we find
\begin{align}
|t_\uparrow|^2&=\frac{[\alpha^2 J+2(h+J)]^2}{4 [\alpha^2 J + (h+J)]^2 +  \left[\frac{(h+J) U^2}{k_F J^2}\right]^2\left[L - \frac{\alpha^2 J+2(h+J)}{\alpha(h+J)}\right]^2}\,,\nonumber\\
|t_\downarrow|^2&=\frac{\alpha^4 J^2}{4 [\alpha^2 J + (h+J)]^2 +  \left[\frac{(h+J) U^2}{k_F J^2}\right]^2\left[L - \frac{\alpha^2 J+2(h+J)}{\alpha(h+J)}\right]^2}\,.
\end{align}
When $L=[\alpha^2 J+2(h+J)]/\alpha(h+J)$, the transmission probabilities are maximized. Likewise, the net spin current, $|t_\uparrow|^2-|t_\downarrow|^2$, is maximized to be $(1+h/J)/[\alpha^2+(1+h/J)]\approx (1+h/J)/\alpha^2$. We plot the probabilities for reflection and transmission in Fig.~\ref{transE0}. Notice that the excitation normally reflected for nearly all values of $L$ except a small range in which the tunneling is peaked.

 \begin{figure}[h]
\includegraphics[width=.5\linewidth]{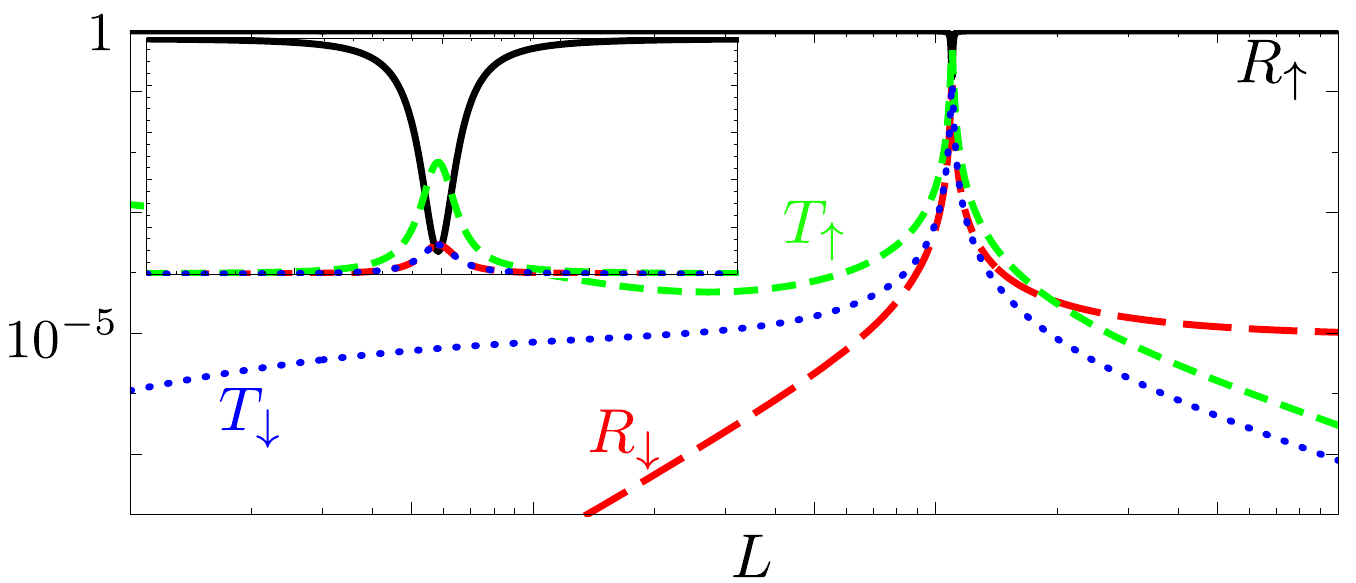}
\caption{Transmission probabilities, $T_\uparrow$ and $T_\downarrow$, and reflection probabilities, $R_\uparrow$ and $R_\downarrow$, as a function of $L$ of an excitation with energy nearly at the gap edge, $E=0.999(h+J)$, and positive spin, $n_\uparrow=1$ and $n_\downarrow=0$. We have taken $\alpha=1$, $h/J=-0.9$, and $h'/J=-0.8$. The main figure is plotted on a log-log scale while the inset is a linear plot on a smaller range of $L$ with the same parameters.}
\label{transE0}
\end{figure}
 \end{widetext}

\end{document}